%% file: main_arxiv.tex
\documentclass[sigconf,10pt,nonacm]{acmart}
\usepackage{amsmath}
\usepackage{enumitem}
\usepackage{xspace}
\usepackage{caption}
\usepackage{subcaption}
\usepackage{float}
\usepackage{cleveref}
\usepackage{algorithm}
\usepackage[noend]{algpseudocode}
\usepackage{mdframed}
\usepackage{listings}
\usepackage{xcolor}

\usepackage{multirow}
\usepackage{pifont}
\usepackage{array, booktabs}
\usepackage{caption}
\usepackage[textsize=tiny,textwidth=0.6in]{todonotes}
\usepackage{adjustbox}
\usepackage[normalem]{ulem} 

\mdfdefinestyle{beautifulcode}{
	backgroundcolor=blue!5, 
	linecolor=blue!50, 
	linewidth=1pt, 
	roundcorner=10pt, 
	innertopmargin=10pt,
	innerbottommargin=10pt,
	innerleftmargin=10pt,
	innerrightmargin=10pt
}

\usepackage{enumitem}

\newcommand{\cmark}{\ding{51}}%
\newcommand{\xmark}{\ding{55}}%

\newcommand{\MyPara}[1]{\vspace{.1em}\noindent\textbf{#1}~}

\usepackage{hyperref}

\begin{document}
	
	\title{Towards Efficient and Practical GPU Multitasking \\in the Era of LLM}
	
	\author{Jiarong Xing$^{\dagger\diamond}$\; 
		Yifan Qiao$^{\dagger}$\;
		Simon Mo$^{\dagger}$\;
		Xingqi Cui$^{\diamond}$\;
		Gur-Eyal Sela$^{\dagger}$\; \\
		Yang Zhou$^{\dagger\ddagger}$\; 
		Joseph Gonzalez$^{\dagger}$\;
		Ion Stoica$^{\dagger}$\; 
		\\[0.5em]
		\emph{$^\dagger$UC Berkeley}\; 
		\emph{$^\diamond$Rice University}\; 
		\emph{$^\ddagger$UC Davis}
	}
	
	\begin{abstract}
		\input{0-abstract}
	\end{abstract}

	\maketitle
	\pagestyle{plain}
	
	\input{1-introduction}

	\input{2-background}
	\input{3-motivation}
	\input{4-design}

	\input{5-discussion}

	\bibliographystyle{ACM-Reference-Format}
	\bibliography{reference}
	
\end{document}

%% file: 0-abstract.tex
GPU singletasking is becoming increasingly inefficient and unsustainable as hardware capabilities grow and workloads diversify. We are now at an inflection point where GPUs must embrace multitasking, much like CPUs did decades ago, to meet the demands of modern AI workloads. 
In this work, we highlight the key requirements for GPU multitasking, examine prior efforts, and discuss why they fall short.
To advance toward efficient and practical GPU multitasking, we envision a resource management layer, analogous to a CPU operating system, to handle various aspects of GPU resource management and sharing.
We outline the challenges and potential solutions, and hope this paper inspires broader community efforts to build the next-generation GPU compute paradigm grounded in multitasking.

%% file: 1-introduction.tex
\section{Introduction}
\label{sec:intro}

\textit{``In today's cloud, we have efficiently virtualized and shared everything---except the most costly resource: GPUs.''}

Despite being used across various tasks, GPUs have historically operated under a \emph{singletasking} paradigm. In this model, one task is given exclusive access to the entire device, aiming to consume as many hardware resources as possible to maximize performance.
However, GPU singletasking becomes increasingly inefficient and unsustainable in today's data centers because of two recent trends.

First, GPU hardware has seen a remarkable advancement in capability. 
Over the past decade, data center GPUs have improved by over 1000$\times$ in peak throughput.
Memory capacity and bandwidth have increased by more than 20$\times$.
Importantly, this growth is expected to continue to meet the growing demands of AI compute~\cite{nvidia-Supercomputer}.
In parallel, AI workloads have become increasingly diverse, with models ranging from 100 million (M) to over 600 billion (B) parameters for different purposes. Their workloads, such as inference request rates, have also become more dynamic, exhibiting significant load fluctuations.

Together, these two trends pose significant GPU utilization challenges in many scenarios. 
For example, small models often fail to saturate large GPUs, GPUs provisioned for peak loads remain idle during off-peak periods, and applications bottlenecked by one type of resource (e.g., memory) leave others underutilized.
As a result, today's data centers often experience low GPU utilization,
e.g., as low as 10\% for inference~\cite{inferencemfu}. 
This inefficiency wastes energy and capital, and will worsen as GPUs continue to grow larger and more powerful.

These limitations have brought GPU systems to an inflection point where adopting a multitasking paradigm---much like CPU systems did decades ago with shared compute and virtualized memory---has become essential to improve hardware utilization and reduce compute costs. 
Achieving this shift requires rethinking how GPUs are managed and shared across tasks. Specifically, we believe effective and practical GPU multitasking should meet at least four key requirements: (1) achieving high resource utilization, (2) providing performance guarantees for performance-critical applications, (3) ensuring fault isolation in multi-tenant environments, and (4) supporting large-scale deployment in data centers.

Recent efforts toward GPU multitasking from both industry~\cite{nvidiaMIG, nvidiaMPS, fractionalGPUs} and academia~\cite{strati2024orion,reef,tgs,ng2023paella,zhang2025improving,zhang2025sgdrc}
share several common limitations that hinder effective and practical GPU multitasking. 
(1) They lack the ability to achieve high resource utilization while providing performance guarantees.
(2) They dedicate great efforts on GPU spatial sharing, which increases scheduling complexity and compromises fault isolation, while overlooking temporal sharing that offers stronger isolation and a cleaner, more transparent implementation.
(3) They primarily focus on GPU compute sharing while overlooking memory sharing, which is increasingly critical as memory consumption grows and becomes more dynamic in large language model (LLM) workloads.
(4) They lack integration with popular orchestration frameworks such as Kubernetes~\cite{kubernetes-web} for large-scale deployment.

To address these limitations, we envision \emph{a GPU resource management layer} that sits between various hardware and applications, functioning as an operating system for GPU multitasking. This layer will manage the shared resources to meet all four key requirements towards efficient and practical GPU multitasking. 

Specifically, we envision it enabling fast, on-the-fly resource partitioning; supporting efficient memory virtualization; cooperatively scheduling resources across applications to achieve high utilization and performance guarantees; ensuring fault isolation; managing network resource sharing; and exposing clean interfaces for large-scale deployment. 
We hope this will reshape the GPU sharing landscape and make efficient, practical multitasking at scale a reality.
Our initial efforts have been included in this open-source ``\texttt{openvgpu}'' project at: \uline{\url{https://github.com/ovg-project.}}

%% file: 2-background.tex
\section{Background and Motivation}
\label{sec:background}

We will primarily use NVIDIA GPUs as an example in the rest of this paper. AMD GPUs have similar designs and customization options through their open-source software stack, ROCm~\cite{amd_rocm}.

\subsection{Background: GPU Hardware/Software}
\label{sec:background:gpu-basic}

\MyPara{Hardware structure.} 
The GPU hardware is designed for massive parallelism, featuring hundreds of Streaming Multiprocessors (SMs) capable of executing thousands of threads concurrently. 
Each SM contains a set of cores, register files, shared memory (L1 cache), and a thread scheduler. 
The threads are grouped into warps
and further organized into thread blocks that the hardware scheduler dynamically assigns to SMs.
GPUs have a multi-tiered memory hierarchy, including the L1 cache within each SM, an L2 cache, and a global memory shared by all SMs, with increased memory size but higher access latency.

\MyPara{Singletasking scheduling.} 
NVIDIA provides CUDA APIs to facilitate GPU programming~\cite{nvidiaCUDAProgramming}. 
CUDA APIs enable developers to write parallel programs as \textit{kernels}, the fundamental unit of GPU programs, akin to functions on CPUs. 
When a kernel is launched, the CUDA runtime prepares it for execution by dividing it into thread blocks. 
The CUDA driver then transfers the kernel and any associated data to GPUs. 
The GPU hardware scheduler handles the final step, dynamically assigning thread blocks to SMs for execution.
CUDA dispatches \emph{one kernel at a time}, with the expectation that \emph{all GPU resources will be dedicated} to executing each kernel as quickly as possible. 
The CUDA task queue enforces \emph{strict sequential}, where kernels are queued and executed in the order they are dispatched.

\subsection{Why Is GPU Singletasking Under Fire?}

Singletasking is well-suited for scenarios where a single application can fully saturate the GPU, such as graphics rendering or batched high-performance computing tasks.
However, with advancements in GPU hardware and the shift toward AI-dominated workloads, singletasking has become increasingly inefficient and unsustainable.

\begin{figure}[t]
  \centering
  \begin{subfigure}[t]{0.49\linewidth}
    \includegraphics[width=\linewidth]{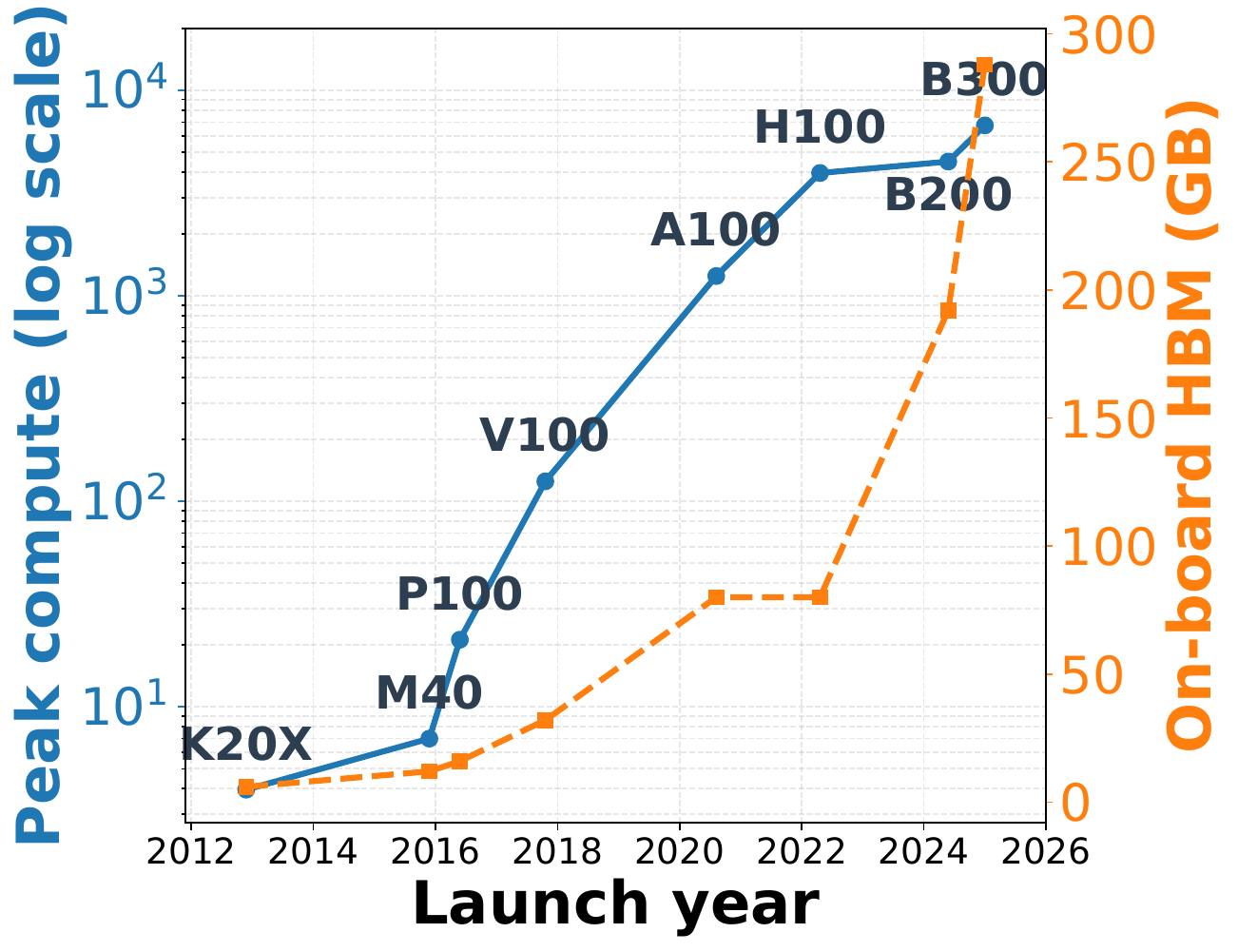}
    \caption{GPU capability leap.}
    \label{fig:gpu-advance}
  \end{subfigure}
  \begin{subfigure}[t]{0.49\linewidth}
    \includegraphics[width=\linewidth]{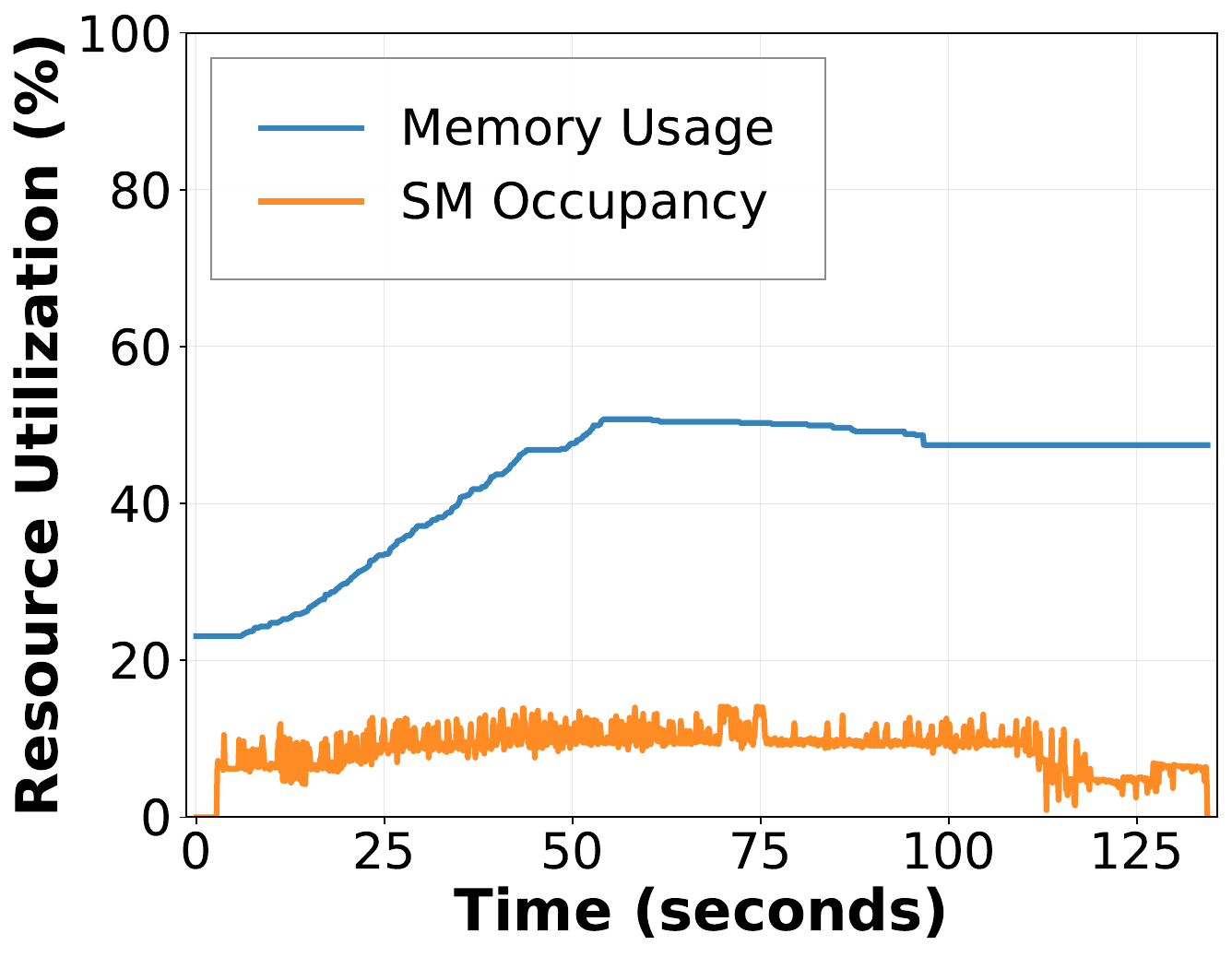}
    \caption{Low GPU utilization.}
    \label{fig:gpu-util}
  \end{subfigure}
  \vspace{-3mm}
  \caption{GPU singletasking is inefficient and unsustainable in the era of LLMs.}
  \vspace{-5mm}
  \label{fig:motivation}
\end{figure}

\MyPara{GPU capability leap.}
Today's GPUs are vastly different from those of a decade ago---the GPU hardware has advanced rapidly. As shown in Figure~\ref{fig:gpu-advance}, the compute throughput of data center GPUs has increased by over 1000$\times$ in the past decade, driven by improved hardware parallelism and support for mixed-precision computation. Meanwhile, GPU memory capacity has expanded by more than 20$\times$, reaching up to 288~GB on today's NVIDIA B300 GPUs. 
In the near future, GPUs are expected to continue this trend~\cite{nvidia-Supercomputer}.

\MyPara{Workload shifts.}
Modern GPU usage is dominated by AI training and inference workloads, which pose new challenges for traditional singletasking paradigms.

\textit{(1) Stretching GPU-to-model size ratio.}
Advancements in AI have produced a range of models for various purposes.
On the one hand, very large models (>600B parameters) continue to emerge~\cite{deepseekv3}, driving the demand for GPUs with increasing capacity. Meanwhile, small distilled models (<10B parameters, e.g., Meta Llama-3.2 (1B, 3B)~\cite{llama31b3b} and gpt-oss-20B~\cite{gpt-oss}) are achieving performance in specialized tasks comparable to large models and getting more deployment~\cite{zhao2024lora, hsieh2023distilling, xiao2024densing}.
This growing GPU-to-model size ratio makes it difficult for small models to fully utilize a GPU's capacity. 
While NVIDIA offers small GPUs, production clusters are typically composed by leading-edge large chips due to their versatility and best flops-to-cost ratio. 

\textit{(2) Increasing load variability.}
AI traffic load can vary drastically over time. 
The various phases of training, such as data loading, 
checkpointing, and network communication, can lead to drastic variability in GPU utilization.
The inference load is even more dynamic. Recent studies~\cite{wang2024towards, qiao2024conserve} show that inference request rates can fluctuate dramatically (e.g., increase by 3$\times$) within minutes. 
This variability is further exacerbated by LLMs, whose resource consumption is highly unpredictable due to its nondeterministic generation process (i.e., auto-regressive decoding)~\cite{pagedattention-vllm@sosp23}.
To ensure quality of service, systems must be provisioned to accommodate peak traffic, which results in resources idle during off-peak periods.

\begin{table*}[t!]
\centering
\resizebox{\linewidth}{!}{
\begin{tabular}{lccccccccccc}
\toprule
\textbf{Features} & \textbf{MIG}~\cite{nvidiaMIG}  & \textbf{FGPU}~\cite{fractionalGPUs} & \textbf{MPS}~\cite{nvidiaMPS} & \textbf{Orion}~\cite{strati2024orion} & \textbf{REEF}~\cite{reef} & \textbf{Paella}~\cite{ng2023paella} & 
\textbf{TGS}~\cite{tgs} & 
\textbf{LithOS}~\cite{coppock2025lithos}  & 
\textbf{BLESS}~\cite{zhang2025improving}  & \textbf{SGDRC}~\cite{zhang2025sgdrc} & \textbf{Ideal}\\
\midrule

\textbf{Target resources}      
    & \textcolor{black}{C(S), M} 
    & \textcolor{black}{M} 
    & \textcolor{black}{C(S)} 
    & \textcolor{black}{C(T,S)} 
    & \textcolor{black}{C(T,S)} 
    & \textcolor{black}{C(T,S)} 
    & \textcolor{black}{C(T),M} 
    & \textcolor{black}{C(T,S)} 
    & \textcolor{black}{C(T,S)} 
    & \textcolor{black}{C(T,S), M} 
    & \textcolor{black}{C(T,S), M} 
    \\

\textbf{High utilization}      
    & \textcolor{red}{\xmark} 
    & \textcolor{red}{\xmark} 
    & \textcolor{red}{\xmark} 
    & \textcolor{green}{\cmark} 
    & \textcolor{green}{\cmark} 
    & \textcolor{green}{\cmark} 
    & \textcolor{green}{\cmark} 
    & \textcolor{green}{\cmark} 
    & \textcolor{green}{\cmark} 
    & \textcolor{green}{\cmark} 
    & \textcolor{green}{\cmark} 
    \\

\textbf{Perf. guarantee} 
    & \textcolor{green}{\cmark} 
    & \textcolor{green}{\cmark} 
    & \textcolor{green}{\cmark} 
    & \textcolor{red}{\xmark} 
    & \textcolor{red}{\xmark} 
    & \textcolor{red}{\xmark} 
    & \textcolor{red}{\xmark} 
    & \textcolor{green}{\cmark} 
    & \textcolor{green}{\cmark} 
    & \textcolor{green}{\cmark} 
    & \textcolor{green}{\cmark} 
    \\

\textbf{Fault isolation} 
    & \textcolor{green}{\cmark} 
    & \textcolor{green}{\cmark} 
    & \textcolor{red}{\xmark} 
    & \textcolor{red}{\xmark} 
    & \textcolor{red}{\xmark} 
    & \textcolor{red}{\xmark} 
    & \textcolor{green}{\cmark} 
    & \textcolor{red}{\xmark} 
    & \textcolor{red}{\xmark} 
    & \textcolor{red}{\xmark} 
    & \textcolor{green}{\cmark} 
    \\

\textbf{Large-scale deploy.} 
    & \textcolor{red}{\xmark} 
    & \textcolor{green}{\cmark} 
    & \textcolor{red}{\xmark} 
    & \textcolor{red}{\xmark} 
    & \textcolor{red}{\xmark} 
    & \textcolor{red}{\xmark} 
    & \textcolor{green}{\cmark} 
    & \textcolor{red}{\xmark} 
    & \textcolor{red}{\xmark} 
    & \textcolor{red}{\xmark} 
    & \textcolor{green}{\cmark} 
    \\
\bottomrule
\end{tabular}
}
\caption{A summary of state-of-the-art GPU multitasking techniques. MIG, FGPU, and MPS are common industry practices, and the rest are from research works.  C: compute; M: memory; S: spatial sharing; T: temporal sharing.}
\vspace{-6mm}
\label{tab:gpu_multitasking}
\end{table*}

\textit{(3) Broadening workload heterogeneity.}
AI workloads are evolving to be more complex and heterogeneous. 
Large model training increasingly relies on small model inference for data preprocessing~\cite{stablediffusion}. 
Compound AI systems~\cite{compound-ai-blog} mix multiple models to complete a job, e.g., a chatbot might use different models to 
generate responses and validate output safety. 
Different models and components have varying resource demands (e.g., compute-bound vs. memory-bound) and performance goals (e.g., interactive vs. batching). Running a single task at a time can easily create bottlenecks in one resource type while leaving others underutilized~\cite{zhu2024nanoflow, zhao2024blendserve}.

\MyPara{Summary: Low utilization due to singletasking.}
As a result, singletasking can easily lead to low GPU utilization. To illustrate this, we measured the compute and memory utilization of a single LLaMA-3-8B model running on an A100 GPU by replaying a production trace from a popular model provider. 
As shown in Figure~\ref{fig:gpu-util}, the GPU remains significantly underutilized for both compute and memory, highlighting the inefficiency of dedicating an entire GPU to a single model in modern serving scenarios.

%% file: 3-motivation.tex
\subsection{From Singletasking to Multitasking}

The above discussion highlights the urgent need for GPU multitasking. This shift would enable more efficient processing of AI workloads on GPUs, reducing capital and energy costs, and further accelerating the growth of AI-driven services. 
GPU multitasking has broad use cases, such as serving multiple models by sharing GPUs, colocating interactive and batching jobs, and optimizing compound AI systems.

\MyPara{Key requirements and challenges.} We argue that efficient and practical GPU multitasking should address the following requirements and challenges:
\begin{itemize}[leftmargin=*, itemsep=0pt, topsep=0pt]
    \item \textit{High utilization.} The primary goal of multitasking is to improve GPU resource utilization, so effective multitasking should maximize overall GPU utilization, including both compute and memory resources.
    This requires the system to reclaim and repurpose unused resources as much as possible to reduce resource idle time.
    
    \item \textit{Performance guarantee.}
    However, high utilization should not come at the cost of application performance, especially for performance-critical tasks, so GPU multitasking must provide \emph{resource isolation} to ensure such tasks receive sufficient resources when needed to meet their performance targets. We argue that the key to resolving the tension between high utilization and performance guarantees lies in \emph{fast and fine-grained, on-the-fly resource partitioning}, which allows the system to dynamically reallocate unused resources and quickly reclaim them as needed.

    \item \textit{Fault isolation}. 
    Moreover, failures or crashes in one application should not affect others running on the same GPU, which is essential for multi-tenant environments. However, current GPU systems do not provide native fault isolation under spatial sharing, as all tasks are forced to share a single execution context (i.e., CUDA context).

    \item \textit{Large-scale deployment}. 
    Last but not least, GPU multitasking should support large-scale deployment in cloud data centers. 
    This requires integration with modern cluster managers (e.g., Kubernetes~\cite{kubernetes-web}) 
    and more recent LLM-specific deployment systems such as LLM-D~\cite{llm-d}, Dynamo~\cite{dynamo}, and OME~\cite{ome}.

\end{itemize}

\subsection{Why Do Existing Techniques Fail?}

Several efforts have made initial progress toward GPU multitasking, as summarized in Table~\ref{tab:gpu_multitasking}. However, upon our examination, they share common limitations that prevent them from meeting the key requirements for efficient and practical GPU multitasking.

\MyPara{Inflexible resource sharing.}
Multi-Instance GPU (MIG) \cite{nvidiaMIG} statically partitions a GPU into independent instances, each with dedicated compute and memory. Fractional GPU (FGPU)~\cite{fractionalGPUs, ailiyun, runai} enables GPU compute sharing via time slicing, combined with statically partitioned GPU memory~\cite{nvidia-plugin}.
Multi-Process Service (MPS)~\cite{nvidiaMPS} supports statically partitioning GPU SM cores into arbitrary sizes for spatial sharing.
however, these industry practices fail to achieve high utilization because their static resource allocations cannot adapt to dynamic workload demands.

\MyPara{No performance guarantees.}
Orion~\cite{strati2024orion}, REEF~\cite{reef}, Paella \cite{ng2023paella}, and TGS~\cite{tgs} improve the flexibility of GPU compute sharing by running kernels from different applications with their own kernel schedulers.
However, different kernels can use arbitrary resources, so they lack the resource isolation needed for performance guarantees.

\MyPara{Lacking fault isolation.}
LithOS~\cite{coppock2025lithos}, BLESS~\cite{zhang2025improving}, and SGDRC~\cite{zhang2025sgdrc}
further enhance resource isolation by enforcing resource limits for each kernel. However, like many other techniques~\cite{strati2024orion,reef,ng2023paella}, their implementations rely on MPS~\cite{nvidiaMPS} for spatial sharing, which compromises fault isolation due to the use of a shared CUDA execution context.

\MyPara{Overlooking memory sharing.}
Most techniques focus on compute sharing while overlooking memory sharing. They often assume memory can be statically partitioned and always fit within GPU capacity. However, this assumption breaks down for LLM workloads, where memory consumption grows dynamically with input and output lengths, making memory more likely to become the bottleneck. This contrasts with traditional AI workloads (e.g., ResNet~\cite{he2016deep}), where memory usage is largely predictable and fixed for a given batch size.

\MyPara{No support for large-scale deployment.}
Although FGPU \cite{fractionalGPUs, ailiyun, runai} and TGS~\cite{tgs} can be integrated with the Kubernetes container orchestration framework~\cite{kubernetes-web}, they lack support for dynamic resource provisioning and do not address network sharing in distributed setups. After a container starts, they all assume this container's resources will remain unchanged, which restricts the flexibility of GPU sharing.

%% file: 4-design.tex
\section{Towards GPU Multitasking}

In this section, we discuss the challenges and potential techniques toward efficient and practical GPU multitasking. We envision these techniques to comprise a unified GPU resource management layer, analogous to a CPU OS, that coordinates compute and memory sharing, achieves high resource utilization while offering performance guarantees, ensures fault isolation, 
and supports deployment at scale.

\subsection{Compute Multiplexing}
\label{sec:comp}

Compute multiplexing is mainly about sharing SM cores, the basic hardware unit that executes GPU kernels in parallel.
There are two ways of sharing SMs: (1) \emph{temporal sharing}, where each task is given all SMs during its time slice, with context switches occurring between tasks over time; and (2) \emph{spatial sharing}, where multiple tasks run concurrently on different subsets of SMs.

\MyPara{Key challenges.}
The key challenge here is to enable flexible compute multiplexing that allows the system to share GPU compute resources at a fine granularity, dynamically adjust partitioning schemes based on workload demands, and enforce resource limits to ensure performance guarantees.

\MyPara{Flexible compute multiplexing.}
For temporal sharing, we argue that kernel-granularity scheduling---intercepting kernel launches and selecting which task's kernel to execute---provides a practical and sufficiently effective solution. Since today's GPUs do not support manual preemption of running kernels, task switching can only occur upon kernel completion. However, modern AI models are often large enough that individual kernel execution times are relatively short compared to the overall request latency, making this approach effective in many cases. For instance, all kernels of a Llama3-8B model can complete within 10ms at a batch size of 8, while end-to-end inference typically takes seconds.

In scenarios where millisecond-level delays matter, we propose leveraging the \emph{time-slicing GPUs} feature available on most modern GPUs~\cite{gpu-time-slice}, which automatically switches execution contexts when a kernel runs out of its allocated time slice. While current implementations of time slicing support only static slice lengths for fair sharing, the recent open-sourcing of GPU drivers from NVIDIA~\cite{open-gpu-kernel-modules} and AMD~\cite{amd_rocm} unlocks the opportunity to implement more flexible and dynamic time-slice scheduling at the driver level.

For spatial sharing, prior works~\cite{coppock2025lithos,zhang2025improving,zhang2025sgdrc} attempt to overcome the inflexibility of MIG~\cite{nvidiaMIG} and MPS~\cite{nvidiaMPS} by constraining the number of SMs a kernel can access through limiting the number of thread blocks it launches. However, this indirect approach often requires modifying kernel implementations, making it impractical. 
Instead, we propose achieving the same purpose by directly controlling the number of SMs a kernel can use at launch time. This can be accomplished using a recently developed library, \texttt{libsmctrl}~\cite{libsmctrl@rtas23}, which enables masking specific SMs accessible to a kernel. This approach operates at the driver level and remains transparent to applications.

\MyPara{Temporal or spatial sharing?}
Temporal and spatial sharing represent different trade-offs. 
Temporal sharing is relatively clean and easier to implement. When realized via time slicing, it offers strong resource isolation but incurs context-switching overhead (e.g., $\sim$100$\mu$s each context switching). 
Spatial sharing, on the other hand, avoids context-switching overhead but introduces additional complexity in scheduling and resource isolation. For example, shared resources like memory bandwidth are difficult to isolate, and SMs cannot be repartitioned while kernels are running due to the lack of manual kernel preemption support.
Moreover, as we will discuss in \S\ref{sec:design:isolation}, spatial sharing will compromise fault isolation due to sharing a single CUDA context, requiring additional mechanisms to mitigate this issue. In contrast, temporal sharing naturally preserves isolation by allowing each task to run in its own context.

There is no definitive reason to favor one strategy over the other in all cases. We argue that the choice between temporal and spatial sharing should depend on workload characteristics and the specific goals being prioritized. If resource and fault isolation are critical, temporal sharing via time slicing offers stronger isolation guarantees. 
In contrast, if maximizing resource utilization is the primary goal, spatial sharing offers the advantage of executing multiple kernels concurrently, which improves hardware efficiency, especially when individual kernels are too small to fully occupy the GPU on their own.

Moreover, certain scenarios allow combining both strategies to achieve temporal–spatial sharing. For example, MIG can be used to spatially partition a GPU among tenants that require strong performance and fault isolation. Within a given tenant, workloads can then temporally share the assigned MIG slices. Recent advances in dynamic MIG slicing~\cite{instaslice-operator, k8s-dra-driver-gpu} further enable tenants to migrate between different slice sizes based on demand, achieving a better balance between resource utilization and performance guarantees.

\subsection{Memory Multiplexing}
\label{sec:mem}

Memory multiplexing has been largely overlooked by prior work, yet it is becoming increasingly critical for two key reasons.
First, memory consumption has grown significantly due to larger model sizes and increased intermediate states (e.g., KV caches) in the LLM era, making memory more likely to become the bottleneck. 
Second, memory usage is now more dynamic and unpredictable due to the non-deterministic nature of LLM generation (i.e., auto-regressive generation), where even identical inputs can yield outputs of varying lengths and thus different memory usage. 
This stands in contrast to traditional neural networks, where memory usage is fixed and predictable given a batch size.

\MyPara{Key challenges.} 
CPU operating systems achieve flexible, fine-grained memory management through virtual memory and page fault handling. However, GPUs have historically been used as singletasking devices and thus lack open interfaces for virtual memory and page fault handling. As a result, conventional GPU applications typically statically reserve a region of physical memory and manually manage it.  In practice, most modern AI applications rely on frameworks like PyTorch~\cite{pytorch}, which abstract away this process and handle GPU memory management behind the scenes.

An emerging opportunity here is the recently released CUDA virtual memory APIs~\cite{nv-vmemapi}, which decouple the allocation of virtual and physical memory, laying the foundation for virtual memory mechanism on GPUs. With these APIs, applications can reserve virtual memory in advance, and the system can map physical memory on demand at a fine-grained page granularity (e.g., 2MB).

However, integrating this virtual memory mechanism with existing AI applications presents new challenges. As mentioned earlier, these applications often implement their own memory management. For example, when memory is released, PyTorch may retain it in their own cache instead of returning it to the system, making it difficult to track actual usage and reclaim unused memory promptly.

\MyPara{Supporting GPU virtual memory.} 
To tackle the challenges of virtual memory without changing application code, we propose intercepting memory-related API calls, such as allocation and deallocation, in the GPU driver. For example, when an application invokes \texttt{cudaMalloc}, we intercept it and replace it with our virtual memory APIs. Later, we could monitor system memory usage to allocate and map physical memory on demand or reclaim it vice versa.

\MyPara{Semantics-aware memory sharing.}
One caveat in applying GPU virtual memory using the above approach is that if applications do not explicitly call memory deallocation APIs (the above PyTorch example), the system lacks the context needed to reclaim that memory.
To bridge this semantic gap, we propose to leverage the customizable memory allocators provided by these frameworks~\cite{pytorchcustomallocator}, and extend them to cooperate with the GPU driver.
This enables the system to accurately track memory usage, promptly reclaim unused memory, and reallocate it to other applications as needed to improve performance.

\MyPara{Memory swapping.}
Out-of-memory errors can occur when GPU memory is oversubscribed, crashing running applications. To mitigate this, we propose a transparent memory swapping technique. When the system detects memory pressure, it automatically evicts pages to CPU DRAM via high-speed GPU-CPU interconnects such as NVLink or PCIe, which are standard on modern GPU servers~\cite{gh200}. To avoid frequent eviction of active pages, 
we can also leverage the semantics-aware memory sharing framework for guided swapping. For example, the PyTorch allocator can share the information about inactive memory with the driver, enabling it to make more informed eviction decisions. 
This approach can effectively mitigate performance degradation caused by memory swapping.

\MyPara{KV cache memory sharing.}
KV cache memory is an exception here, as its regular usage pattern actually simplifies the problem. KV cache is commonly managed by today's open-source LLM inference engines, such as vLLM~\cite{pagedattention-vllm@sosp23} and SGLang~\cite{sglang@nips24}. Although current implementations statically reserve GPU physical memory without considering memory sharing, they provide well-defined interfaces for KV cache management (e.g., alloc/free functions). 
This offers an easier path to enable memory sharing---as proposed in our prior work~\cite{yu2025prism}, replacing their KV cache management functions with our virtual memory implementation enables flexible memory sharing without modifying the GPU driver.

\subsection{Resource Coordination}
\label{sec:design:scheduler}

The techniques in \S\ref{sec:comp} and \S\ref{sec:mem} offer basic primitives for GPU multitasking. To achieve high utilization and performance guarantees, we also need effective resource coordination.

\MyPara{Elastic resource partitioning.}
We propose to provision resources in two types: \emph{guaranteed} and \emph{preemptible}.
A task is always entitled to uninterrupted use of its guaranteed resources. When spare GPU capacity exists, the task can utilize it as preemptible resources to enhance performance. However, these preemptible resources can be reclaimed by the system to fulfill guaranteed resource requirements of other processes. Meanwhile, if a task does not fully utilize its guaranteed allocation, we may temporarily allocate the unused portion as preemptible resources for other tasks until the original task requires it.

\MyPara{Cooperative preemption.}
When a task requires guaranteed resources currently used by preemptible tasks, the system must reclaim those resources promptly to ensure performance. 
Reclaiming compute resources is relatively straightforward: the system can wait for the current time slice or kernel to complete and then readjust the allocation.
In contrast, memory preemption is more challenging because the system must preserve application data---na\"ively reclaiming memory by deleting data can lead to application crashes. 
To address this, we can reuse the transparent memory swapping mechanism discussed in \S\ref{sec:mem}. Instead of deleting data, the system can temporarily swap it to CPU memory. Additionally, by involving the extended PyTorch memory allocator, we can gain application-level semantics to help guide swapping decisions. This enables the system to prioritize evicting unused or inactive memory and avoid frequent swapping of active data, which would otherwise degrade performance.

\begin{figure}[t]
    \centering
    \includegraphics[width=0.7\linewidth]{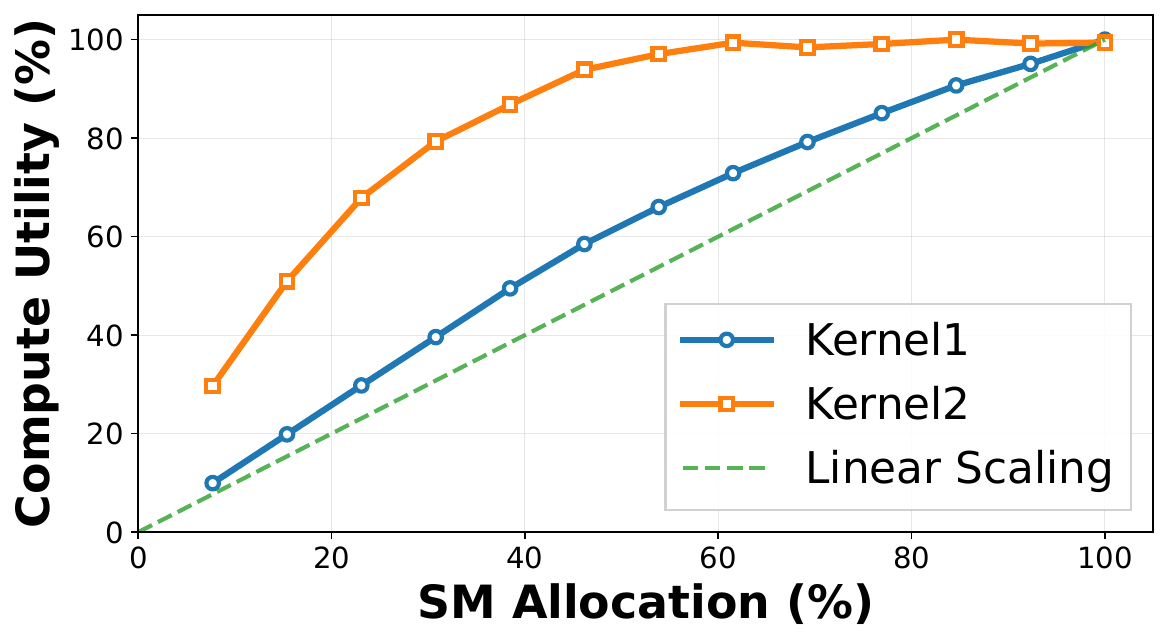}
    \vspace{-3mm}
    \caption{The compute utility of two kernels.}
    \vspace{-6mm}
    \label{fig:utility-curve}
\end{figure}

\MyPara{Utility-guided coordination.}
The available preemptible resources can be shared across tasks using customizable policies such as max-min fair scheduling~\cite{maxminfair}. Here, we propose a utility-guided policy that dynamically allocates resources based on task efficiency, improving overall hardware compute utilization.
The key insight is that different kernels yield different performance gains as more resources are provisioned. To quantify that, we define a metric called \emph{utility}, which quantifies the fraction of performance a kernel achieves relative to using the full GPU. 
As illustrated in Figure~\ref{fig:utility-curve}, Kernel1 exhibits a near-linear utility increase, whereas Kernel2 quickly saturates. This means that allocating the same additional resources can lead to varying utility improvements across kernels. For example, after 60\%, increasing SM allocation for Kernel2 will only return marginal benefits.
The system can leverage these utility curves to guide resource allocation and optimize the overall utility of preemptible resources.

\subsection{Fault Isolation}
\label{sec:design:isolation}

Fault isolation is critical for GPU multitasking in multi-tenant environments. It ensures that a fault in one task, whether triggered by compute units, memory controllers, or other components, will not compromise the correctness or execution of other tasks sharing the same GPU.

On GPUs, each process maintains a CUDA context that encapsulates its execution state. This context is essential for correct execution; if an error occurs within a context, the associated process will crash. GPU temporal sharing naturally provides fault isolation, as each task runs in its own process with an independent context. As a result, faults in one task do not propagate to others.

However, for spatial sharing, fault isolation is only supported through MIG~\cite{nvidiaMIG}, which physically partitions the GPU hardware into static, fixed-size slices. Without MIG, the only way to spatially share the GPU is through MPS~\cite{nvidiaMIG}, which merges multiple processes into a single shared CUDA context. As a result, MPS-based solutions compromise fault isolation, as errors in one task can propagate to the entire shared context and crash everything.

We argue that this limitation stems more from the GPU software stack than from fundamental hardware constraints. As reported by NVIDIA~\cite{nvidiaMIG}, modern GPUs already include multiple compute engines, memory controllers, and copy engines that can be assigned independently to different processes. MIG relies on this capability to provide isolation through static partitioning.
As a result, we propose rearchitecting the GPU runtime and driver to support fault isolation as a first-class feature under spatial sharing. 
Our key insight is to track the specific hardware components (e.g., which SM group) allocated to each kernel at launch time in the GPU kernel drivers~\cite{open-gpu-kernel-modules}. When a hardware fault occurs, the driver can identify the faulting kernel and process based on the hardware component allocation, terminate it gracefully, and allow other tasks to continue running without disruption.

\subsection{Large-Scale Deployment in Cloud}

AI workloads are predominantly hosted in large-scale cloud data centers. However, scaling GPU multitasking to multi-GPU servers and multi-node clusters introduces new challenges in instance management and network sharing.

\MyPara{Integration with management tools.}
Kubernetes~\cite{kubernetes-web} has become the de facto container orchestration framework in cloud data centers, 
and recent years have seen the emergence of LLM deployment frameworks such as LLM-D~\cite{llm-d}, Dynamo~\cite{dynamo}, and OME~\cite{ome} that build on top of it.
To support large-scale deployment, GPU multitasking must be seamlessly integrated into their ecosystems.
However, the current Kubernetes' device-plugin model assumes static devices: GPUs (or MIG slices) are enumerated at \texttt{kubelet} startup, and the scheduler then treats them as immutable, integer-count resources. Thus, dynamic GPU sharing will violate these assumptions, make it incompatible with Kubernetes.
Fortunately, recent proposals for Dynamic Resource Allocation (DRA) in Kubernetes~\cite{kubernetes-dra} offer a promising path forward. By extending DRA, we can potentially enable flexible, fine-grained GPU sharing at data center scale with native container orchestration support.

In addition, these frameworks already include their own control loops, autoscalers, and routing layers, so the multitasking system should also coordinate with their scheduling and scaling decisions.
It may also need to manage KV cache tiering across memory hierarchies, and expose standardized metrics for joint autoscaling and routing.
To maintain SLO attainment at scale, workload-aware placement, dynamic adaptation, and failure-safe preemption are equally critical~\cite{yu2025prism}.

\MyPara{Network sharing.}
Tasks running on the same GPU may also share network resources. This is relatively straightforward with temporal sharing, as each task can use the network as usual during its time slice. However, spatial sharing incurs several challenges.
First, the system must enable the communication library like NCCL~\cite{nccl-web} to recognize GPU slices as independent devices. In current implementations, NCCL treats each physical GPU as a single device, unless MIG~\cite{nvidiaMIG} is enabled. One possible solution is to expose logical GPUs by extending kernel-level support, analogous to how SR-IOV~\cite{network-virtual-sriov-hpca10} creates virtual functions for network devices. This would allow each GPU slice to be addressed and scheduled independently by the communication stack. 

The second challenge is isolating communication resources to provide performance guarantees. On modern GPU machines, intra-node communication typically uses NVLink~\cite{nvlink-web}, while inter-node communication is efficiently handled via GPUDirect RDMA~\cite{nvidiaGPUDirect-web}. Communication libraries such as NCCL~\cite{nccl-web} can automatically choose between these paths. 
However, current hardware lacks native support for slicing or partitioning NVLink bandwidth.
A potential workaround is to intercept the communication kernels launched by NCCL (for both NVLink and RDMA transfer) and enforce custom scheduling or sharing policies at that level.

%% file: 5-discussion.tex
\section{Open Problems}

\MyPara{Security isolation.}
GPU multitasking should ensure security, e.g., protecting legitimate tasks from malicious or buggy co-running tasks. Our proposed virtual memory design enforces separate virtual address spaces for different tasks, providing basic memory safety. However, it remains as an open problem to fully eliminate other security vulnerabilities, such as side-channel attacks.

\MyPara{Memory bandwidth control.}
Memory bandwidth isolation is also important for performance guarantees under GPU spatial sharing. 
However, current GPU hardware does not provide mechanisms for directly controlling memory bandwidth. Even MIG~\cite{nvidiaMIG}, which physically partitions other GPU resources, fails to provide accurate control over memory bandwidth.
A key challenge is that memory bandwidth consumption tends to not scale linearly with the number of active SMs. For example, just 20\% of SMs can potentially saturate the entire memory bandwidth.
One potential workaround could be to instrument memory load/store instructions and insert no-ops to delay execution. Tuning these delays allows the system to throttle memory access and control bandwidth.